\documentclass[useAMS,usenatbib]{mn2e}

\usepackage[pdftex,pdfpagemode={UseOutlines},bookmarks,bookmarksopen,colorlinks,linkcolor={blue},citecolor={green},urlcolor={red}]{hyperref}
\usepackage{journals_macros}
\usepackage{graphicx}
\usepackage{natbib}
\usepackage{xspace}
\usepackage{amsmath}

\usepackage{times}

\def\mhmpc{\,h^{-1}~\rm{Mpc}}

\def\mhmpcc{\,h^{-3}~\rm{Mpc^3}}
\def\mhgpcc{\,h^{-3}~\rm{Gpc^3}}

\def\mhmsun{\,h^{-1}~\rm{M_{\odot}}}
\newcommand{\hmpc}{$\,h^{-1}$ Mpc\xspace}

\title[Reconstructing the halo distribution below the resolution
  limit] {Reconstructing the distribution of haloes and mock galaxies
  below the resolution limit in cosmological simulations}

\author[S. de la Torre \& J. A. Peacock]
 {Sylvain de la Torre\thanks{E-mail: sdlt@roe.ac.uk}
  and John A. Peacock \\
  SUPA\thanks{Scottish Universities Physics Alliance}, Institute for Astronomy, University of Edinburgh, Royal Observatory, Blackford Hill, Edinburgh EH9 3HJ, UK}

\begin{document}

\date{Accepted 2013 July 18.  Received 2013 June 3; in original form 2012 December 14}

\pagerange{\pageref{firstpage}--\pageref{lastpage}} \pubyear{2012}

\maketitle

\label{firstpage}

\begin{abstract}
\noindent We present a method for populating dark matter simulations
with haloes of mass below the resolution limit. It is based on
stochastically sampling a field derived from the density field of the
halo catalogue, using constraints from the conditional halo mass
function $n(m|\delta)$. We test the accuracy of the method and show
its application in the context of building mock galaxy samples. We
find that this technique allows precise reproduction of the two-point
statistics of galaxies in mock samples constructed with this
method. Our results demonstrate that the main information content of a
simulation can be communicated efficiently using only a catalogue of
the more massive haloes.

\end{abstract}


\section{Introduction}

The distribution of luminous matter in the Universe is a rich source
of information concerning the large-scale mass distribution, the
formation and evolution of baryonic structures, plus the overall
properties of the Universe. In this way, future large galaxy surveys
will yield extremely precise measurements of quantities such as the
global expansion history and growth rate of structure from
measurements of the galaxy power spectrum
\citep[e.g.][]{laureijs11,schlegel11}.  But achieving high statistical
precision is only possible if we have a full understanding of all the
potential systematic biases arising from survey selection -- and
realistic surveys are sufficiently complex that the necessary
calibration of statistical methods can only be achieved by using mock
survey realisations.  Moreover, mock datasets represent one of the
most efficient ways of estimating uncertainties related to the
statistics of interest, including both estimation and sample variance
errors; typically 100-1000 independent realisations are required for
this purpose.  Naturally, robust answers to these questions require
the mocks to be as realistic as possible, although they do not need to
match reality in every respect, so long as they contain the main
complicating factors that could be a source of error in the real data.

For these reasons, dark matter N-body simulations coupled with
semi-analytical treatments of galaxy formation or halo occupation
distribution (HOD) techniques have become standard methods for
producing mock survey samples. These two-layered methods reflect our
current understanding of cosmology and galaxy formation, where the
large-scale structure of the Universe, dominated by dark matter,
evolves through gravity; galaxies then form inside the dark matter
haloes by the collapse and cooling of baryonic gas into the potential
wells that they provide. However, dark matter N-body simulations are
finite and their usefulness is limited by the volume and mass
resolution that they can probe. For cosmological surveys, the probed
volume directly influences the statistical error with which one can
measure the cosmological parameters. Conversely, mass resolution is
more important for galaxy evolution surveys, in which one is more
interested in a complete census of galaxies, i.e. probing the whole
range of galaxy masses and associated physical properties. These
competing criteria of volume and resolution mean that all existing
simulations are a compromise.

One way of tackling this limitation is to make an approximate
reconstruction of the distribution of the omitted low-mass haloes.
Given predictions of the halo bias and abundance at the lowest masses,
one can aim to recover the missing information below the resolution
limit. In principle, a good deal of information is available for this
task, since about half of the mass in most simulations is not resolved
into haloes. But for large simulation volumes, handling this quantity
of particle data is cumbersome, and often only the catalogue of
resolved haloes can be efficiently transmitted. Our aim here is thus
to see to what extent the distribution of all haloes can be inferred
from only those that are detected in a simulation. We present in this
paper a method of this sort, which we show to be particularly
effective in the context of building mock galaxy surveys. The idea of
synthesizing a halo population from a smooth density field has been
explored in the past \citep{scoccimarro02,angulophd}, but the novelty
here is to show that this can be done directly from a catalogue of the
massive haloes. The aim of this paper is to provide a general
framework that allows the creation of mock halo and galaxy catalogues
with percent-level accuracy on the two-point statistics. We do not
consider higher-order statistics here, but the method could
potentially be extended to deal with these.

\newpage

\section{Method}

The proposed method consists of two steps: one first uses the
simulation halo catalogue to estimate a halo density field which is
then sampled stochastically to obtain haloes with mass below the
resolution limit and with the correct abundance and bias. To predict
the number of haloes of different masses in each region of the
simulation, i.e. the conditional halo mass function, we use the
peak-background split formalism \citep{bardeen86,cole89}. The shape
the halo mass function $n(m)$ and bias factor $b(m)$, which are the
basic ingredients entering in the conditional halo mass function, have
to be extrapolated to the masses below the nominal minimum halo mass
of the simulation or to be assumed from theory. In the following
subsections we describe in detail the two parts of the method.

\subsection{Halo density field estimation}

The main idea is to use the simulation halo catalogue, which
preferentially traces the densest environments, to infer the full
range of overdensity. The first step is to estimate the halo density
field traced by the haloes originally present in the simulation. There
are several ways to estimate the density field, the simplest being to
count the number of objects on a cubical grid, assigning each halo
to the closest grid node. Generally the means of assigning objects to
grid nodes and the grid size have an impact on the accuracy of the
recovered density field. Optimally, we would like to use cells as
small as possible, so as to probe the smallest-scale density
fluctuations, but also large enough to avoid introducing shot
noise. The optimal grid size will then depend on the number density of
haloes in the simulation and, in turn, on the nominal halo mass
resolution. One way of reducing the shot noise in the reconstructed
density field is to use Delaunay tessellation. In that case, instead
of using fixed-size cells to estimate number densities, one uses
tetahedra whose size varies adaptively depending on the local number
of objects. The resulting density field estimates can then be
interpolated onto a fine grid for convenience. This method allows the
reduction of the shot noise contribution, while retaining
high-resolution information when it is available. Other adaptive
smoothing methods based on e.g. nearest neighbours could also be
used. We show in the next section the improvement on the halo density
field estimation that this can produce.

\subsection{Low-mass halo population} \label{sec:population}

Once a continuous halo density field is estimated, one can use the
expected number of haloes of mass $m$ in each cell of mass overdensity
$\delta$, i.e. the conditional halo mass function $n(m|\delta)$, to
populate the simulation with haloes of mass below the resolution
limit. The halo density field $\delta_h$ is biased with respect to the
mass density field $\delta$ and consequently has to be de-biased prior
to being used to predict the number of expected low-mass haloes.

We follow the peak-background split formalism and write the
conditional halo mass function as
\begin{equation}
n(m|\delta)=n(m)(1+\left<\delta_h(m)|\delta\right>), \label{nmd}
\end{equation}
where $n(m)$ is the (unconditional) halo mass function and
$\left<\delta_h(m)|\delta\right>$ is the function describing the
biasing of haloes of mass $m$. In the case of sufficiently large
cells, density fluctuations become linear and we can assume
$\delta_h=b(m)\delta$. In this limit Equation (\ref{nmd}) simplifies
to
\begin{equation}
n(m|\delta)=n(m)(1+b(m)\delta) \label{nmd1}
\end{equation}
where $b(m)$ is the large-scale linear halo bias factor. In practice,
$n(m)$ and $b(m)$ have to be specified for mass values below $m_{\rm
  lim}$, the minimum halo mass of the simulation. For this one can
either use analytical forms, extrapolate these functions in the
simulation itself or use higher-resolution simulations. The
extrapolation is relatively straightforward because those functions
show only weak and relatively easily predictable variations with halo
mass in the low-mass regime.

Equation (\ref{nmd1}) is valid for densities estimated on large scales
where non-linear fluctuations are smeared out. However we would like
to have a model that accounts to some extent for bias non-linearities
which are present on small scales. One simple (local) non-linear
biasing model that we can use is the power-law bias model
\citep[e.g.][]{mann98,narayanan00} for which the halo bias is defined
as
\begin{equation}
1+\delta_h\propto (1+\delta)^{b(m)}. \label{nmd2}
\end{equation}
This model has a certain number of advantages: it naturally avoids
negative densities and depends only on one parameter. Furthermore such
a power-law model has empirical support to the extent that it gives a
good match to the relative biasing of different classes of galaxies
\citep{wild05}. We will show in Section \ref{sec:test} that it is
accurate enough for the purpose of the present method. We use a purely
deterministic bias prescription here, although it is clear empirically
\citep{wild05} that a small stochastic component should be included in
Equation \ref{nmd2}.  As shown below, however, sufficiently accurate
results are obtained at the two-point level without this extra
complication being required.  While using the power-law bias model in
Equation (\ref{nmd}), one obtains a conditional halo mass function of
the form
\begin{equation}
n(m|\delta)\propto n(m)(1+\delta)^{b(m)}. \label{nmd3}
\end{equation}
Because the halo density field is biased and the mass overdensity that
enters in Equation (\ref{nmd3}) is unknown \emph{a priori}, one has to
rewrite the conditional mass function in terms of the halo overdensity
$\delta_h$. If we assume the same biasing model to de-bias the
original halo density field, then the final conditional halo mass
function that we can use to populate the simulation in low-mass haloes
is
\begin{equation}
n(m|\delta_h)\propto n(m)(1+\delta_h)^{b(m)/b_0}, \label{nmd4}
\end{equation}
where $b_0$ is the effective bias of the original halo population,
defined as
\begin{equation}
b_0=\frac{\int_{m_{\rm lim}}^\infty b(m)n(m)~dm}{\int_{m_{\rm lim}}^\infty n(m)~dm}.
\end{equation}
Note that there would be a factor $m$ inside the integral if we had
chosen to weight haloes by mass. But number weighting reduces both
non-linear bias and shot noise from finite numbers of haloes. In
practice the normalisation of Equation (\ref{nmd4}) is imposed
empirically by requiring $\left<\delta_h\right>=0$ when volume
averaging over all cells of the simulation. Finally, the number of
low-mass haloes in each cell is randomly drawn by Poisson sampling the
$n(m|\delta_h)$ and these haloes are spatially distributed in a
uniform fashion within the cell. With this procedure, the low-mass
haloes will not exhibit any clustering on scales below the size of the
cells, but again we show below that this has no impact on the
precision of our results. The unresolved haloes are only a
perturbation to the signal from the resolved haloes and we are
interested in the total signal, rather than the low-mass haloes in
isolation.

One could of course have used the mass density field or an estimate of
it from dark matter particles in the simulation and worked directly
from the mass density field $\delta$ using Equation (\ref{nmd3}), but
using the halo catalogue allows the reconstruction method to be
applied to public simulation datasets where the full particle density
field is typically not made available.

\section{Tests on simulation data} \label{sec:test}

\begin{figure*}
\includegraphics[width=176mm]{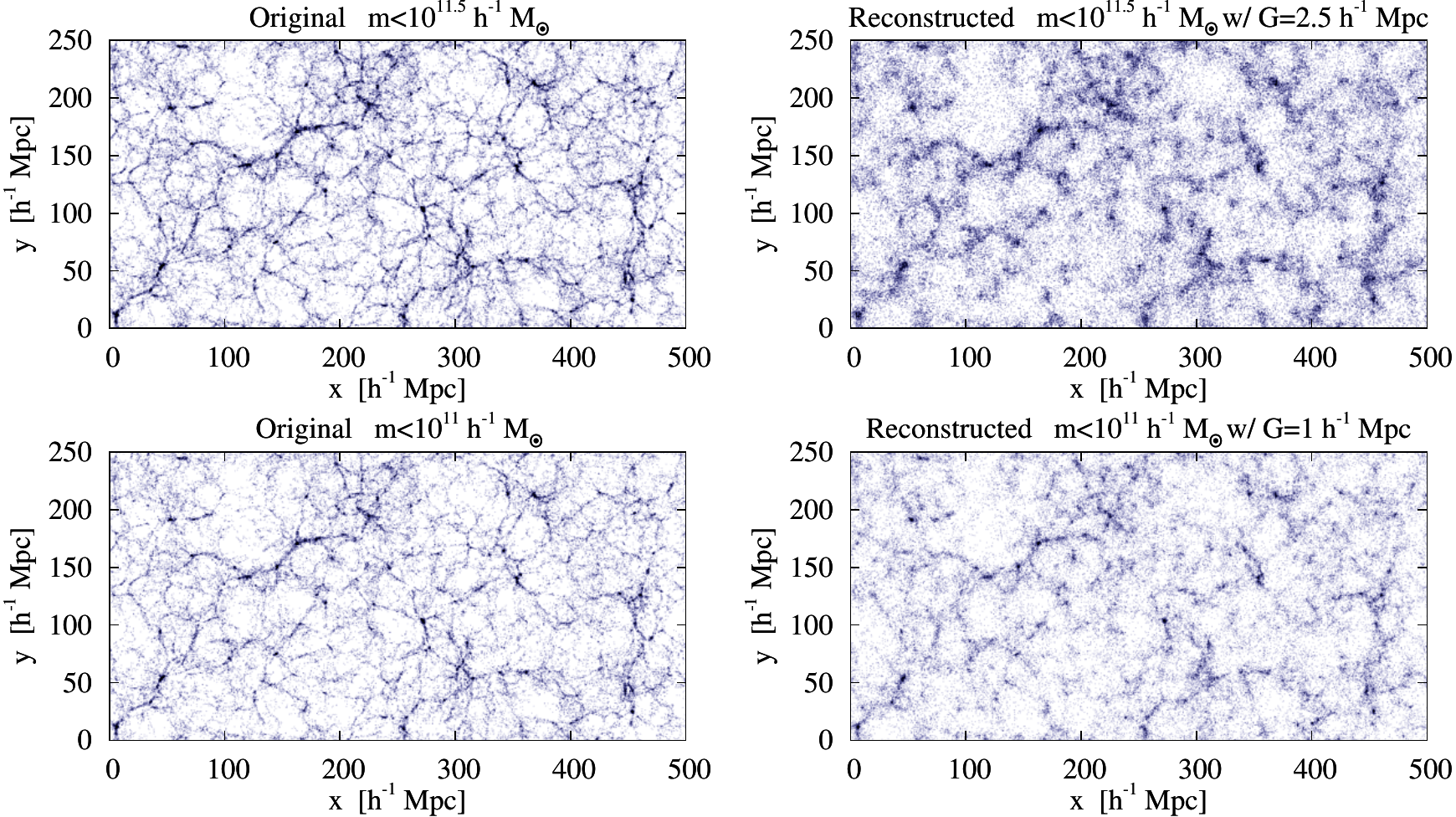}
\caption{Comparison of the continuous density fields of original (left
  panels) and reconstructed haloes (right panels) in a slice of $500
  \times 250 \times 15\mhmpcc$ from the Millennium simulation, for two
  cuts in halo mass corresponding to $m<10^{11.5}\mhmsun$ (top panels)
  and $m<10^{11}\mhmsun$ (bottom panels). In the $m<10^{11.5}\mhmsun$
  case, the reconstruction used a grid of size $G=2.5\mhmpc$, while in
  the $m<10^{11}\mhmsun$ case, a grid of size $G=1\mhmpc$ was used.}
\label{fig1}
\end{figure*}  

We test the reconstruction method on the Millennium simulation
\citep{springel05} which probes a volume of $0.125\mhgpcc$ with a mass
resolution of $m_p=8.6\times10^8\mhmsun$ in a $\Lambda\rm{CDM}$
cosmology with $(\Omega_m,~\Omega_\Lambda,~\Omega_b,~h,~n,~\sigma_8) =
(0.25,~0.75,~0.045,~0.7,~1.0,~0.9)$. We will also make use in the
following of the MultiDark Run 1 (MDR1) dark matter N-body simulation
\citep{prada12}. MDR1 probes a larger volume of $1\mhgpcc$ with a mass
resolution of $m_p=8.721\times10^9\mhmsun$ in a $\Lambda {\rm CDM}$
cosmology with $(\Omega_m,~\Omega_\Lambda,~\Omega_b,~h,~n,~\sigma_8) =
(0.27,~0.73,~0.0469,~0.7,~0.95,~0.82)$. In both simulations, the dark
matter haloes have been identified from the dark matter particle
distribution using a friends-of-friends algorithm and we use only the
haloes identified in the snapshots at $z=0.1$. The minimum halo mass
in the Millennium and MultiDark halo catalogues are respectively
$m_{\rm lim}=10^{10.5}\mhmsun$ and $m_{\rm lim}=10^{11.5}\mhmsun$.

\begin{figure}
\centering
\includegraphics[width=88mm]{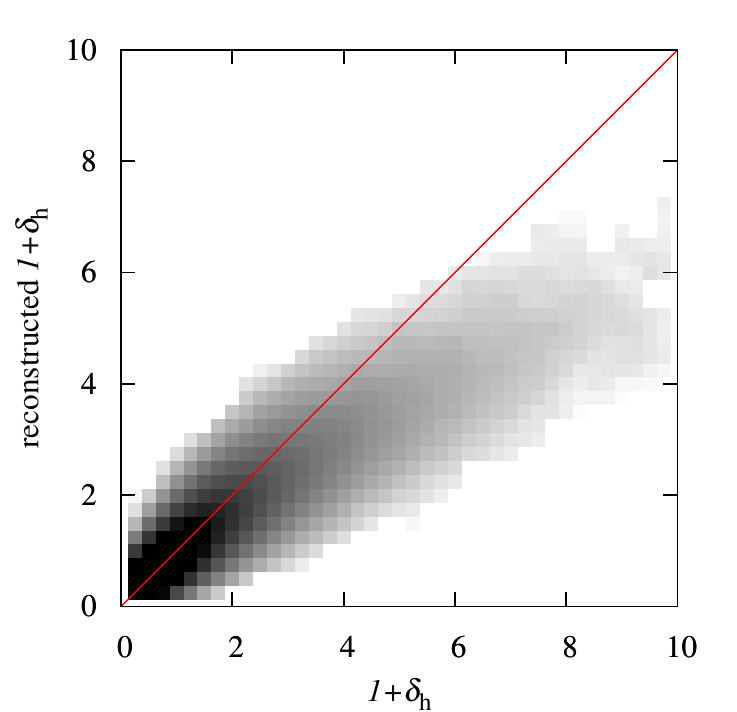}
\caption{Comparison of the densities of original and reconstructed
  halo fields in the Millennium simulation for haloes with
  $m<10^{11.5}\mhmsun$. The reconstruction used a grid of size
  $G=2.5\mhmpc$. The halo overdensities $\delta_h$ have been estimated
  in cubical cells of $5^3\mhmpcc$. The line shows the case where
  original and reconstructed halo densities are equal.}
\label{fig4}
\end{figure}

We estimate the halo density field by measuring the halo density
contrast defined as
$\delta_h(\bmath{r})=(N(\bmath{r})-\left<N\right>)(\left<N\right>)$
where $N(\bmath{r})$ and $\left<N\right>$ are respectively the number
of haloes in a cell centred at position $\bmath{r}$ and the mean
number of haloes per cell. Given the halo number density, the optimal
choice of cell size falls between 2.5\hmpc and 5\hmpc, in order to
have a few haloes per cell on average. We choose a grid size of
$G=2.5\mhmpc$ and estimate the halo density field using different
methods: the grid-based method with Nearest Grid Point (NGP) and
Cloud-In-Cell (CIC) assignment schemes and the Delaunay Tessellation
(DT) method. We choose haloes above a limit between $10^{10}$ and
$10^{11.5}~\mhmsun$ and reconstruct the smaller haloes using the
conditional mass function of Equation (\ref{nmd4}). In this test, we
assumed for $b(m)$ and $n(m)$ the forms calibrated on N-body
simulations by \citet{tinker08} and \citet{tinker10}. The output of
the reconstruction is illustrated in Fig. \ref{fig1}, which shows the
spatial distribution of original and reconstructed haloes in a thin
slice of the Millennium simulation.  A more quantitative comparison
between the original and reconstructed fields is shown in
Fig. \ref{fig4} where the densities are compared cell-by-cell. The
agreement is good up to $1+\delta\simeq 3$, although beyond this the
reconstruction falls below the true value.  Some underestimation of
high-density knots is probably inevitable with a finite resolution,
and could be corrected with a more complex nonlinear halo bias
relation. This is not necessary for two-point applications, since the
fraction of haloes in such extreme overdensities is small, but there
could well be an impact on higher-order statistics.

\begin{figure}
\centering
\includegraphics[width=84mm]{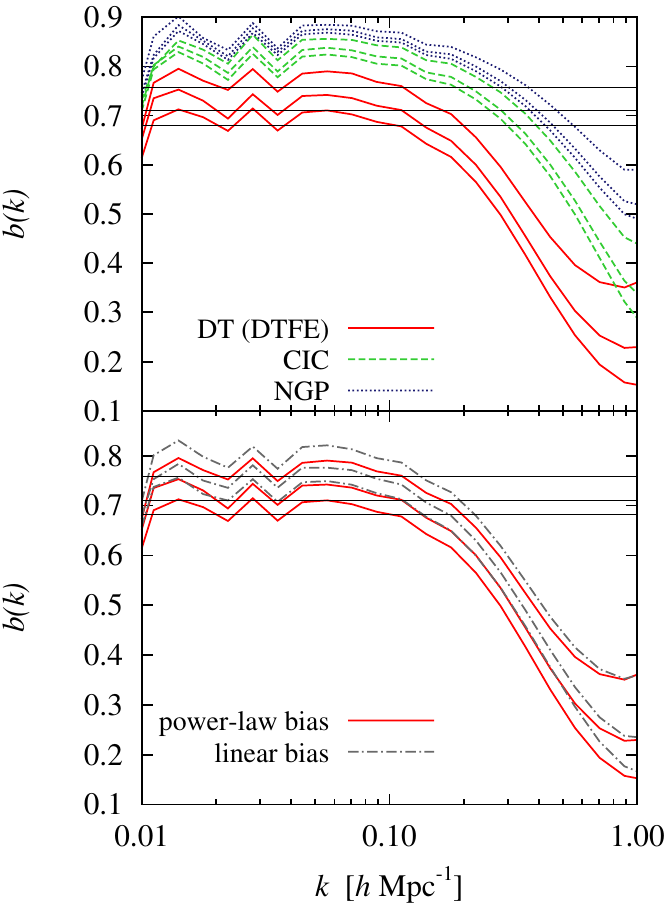}
\caption{Top: scale-dependent bias of the reconstructed haloes
  obtained using different methods applied to the MultiDark data to
  estimate a continuous halo density field. For each of the methods:
  DT (dotted), CIC (dashed), and NGP (dot-dashed), the three curves
  correspond to samples of haloes with mass in the ranges
  $10^{10}<m<10^{10.5}\mhmsun$, $10^{10.5}<m<10^{11}\mhmsun$, and
  $10^{11}<m<10^{11.5}\mhmsun$, respectively from top to bottom,
  reconstructed using the haloes above the upper mass limit in each
  case. Bottom: scale-dependent bias of the reconstructed haloes
  obtained using linear and power-law bias prescriptions in the
  reconstruction. In these cases, the DT method has been used to
  estimate a continuous halo density field.  In all panels, the
  horizontal lines are the linear halo bias predictions by
  \citet{tinker10} for the three mass bins considered.}
\label{fig2}
\end{figure}

To test the accuracy of the method we perform the reconstruction on
the MultiDark simulation, which gives us a better probe of the
large-scale halo clustering. We measure the halo bias in the low-mass
regime from the reconstructed halo catalogue. The halo bias has been
estimated by first measuring the halo power spectrum $P(k)$ and then
taking the square root of the ratio between the halo power spectrum
and that of mass. In this, we assumed the non-linear mass power
spectrum given by CosmicEmu \citep{lawrence10}.

The recovered halo biases in mass bins below the resolution limit are
shown in Fig. \ref{fig2}, which compares the results of using
different estimates of the halo density field as well as different
biasing models. In this figure, the measured halo bias is shown as a
function of the wavenumber for the three mass bins:
$10^{10}<m<10^{10.5}\mhmsun$, $10^{10.5}<m<10^{11}\mhmsun$, and
$10^{11}<m<10^{11.5}\mhmsun$. We find that the DT method as
implemented in the DTFE code \citep{cautun11} provides better results
than the grid-based estimator with CIC and NGP assignment schemes. The
large-scale bias, expected to asymptote to linear theory predictions,
is in very good agreement with the predictions of \citet{tinker10} in
the case of DT, whereas for the other methods the bias is clearly
overestimated. This is particularly true in the case of NGP. The DT
method better accounts for local variations in number density,
reducing the shot noise in the reconstruction and giving a better
sampling of the most extreme environments. In the NGP and CIC cases,
the significant shot-noise contributions translate into additional
scale-dependent components in the power spectrum and estimated
bias. In this exercise, we pushed the methods towards their limits by
considering a very small grid size of $2.5\mhmpc$. However, if we
increase the grid size to $5-10\mhmpc$, the recovered halo biases come
to agreement and we find that the three methods converge to the same
values.

The biasing scheme that enters in the conditional mass function has
also some impact on the recovered halo clustering, in particular for
small grid size density field reconstruction such as the one
considered here. We show in the bottom panel of Fig. \ref{fig2} the
effect on the recovered halo bias when assuming a linear or power-law
bias model as describe in Section \ref{sec:population}. In both cases
we use the halo density field reconstructed with the DT method. We
find that the linear model tends to overestimate the large-scale
linear bias for low-mass haloes compared to the power-law model, which
instead allows us to recover the linear bias predictions of
\citet{tinker10} at the few percent level.

It is noticeable in Fig. \ref{fig2} that, as is inevitable, one cannot
reconstruct the highest $k$ regime of the halo power spectrum. This
however does not really matter for the purpose of galaxy mock
construction, since the overall galaxy power spectrum is dominated by
the 1-halo term in this regime, as we will show in the next section.

Another aspect which can be important for creating realistic halo
catalogues is the assignment of velocities to the newly created
haloes. Their velocity should sample the underlying velocity field
which can be estimated from the original haloes. The velocity field is
a volume-weighted quantity and for this reason it is more difficult to
measure than the density field from the set of original haloes. It has
been shown that the DT method is particularly efficient at recovering
the velocity field and it naturally avoids the velocity field to be
artificially set to zero in regions where there are no haloes, which
can be the case for mass-weighted approaches based for instance on
interpolating the velocities to a grid \citep[e.g.][]{jennings12}. The
estimated velocity field with DT can thus be used and interpolated at
the position of the reconstructed haloes so that reliable velocities
can be assigned to them.

\section{Application to galaxy mock sample construction} \label{sec:mock}

The reconstruction of the halo density field below the resolution
limit in cosmological simulations is particularly useful in the
context of building realistic galaxy mock surveys. As explained
earlier, forthcoming large cosmological surveys of galaxies will need
a large number of mock survey realisations, and we need these mocks to
include galaxies of very low luminosity/stellar mass. These dim
galaxies sit in low-mass haloes, so that a method such as the present
one is required to restore such missing haloes. In the following, we
apply our halo reconstruction method and test its efficiency in this
context.

An efficient way to build galaxy mock samples is to use the Halo
Occupation Distribution (HOD) formalism
\citep{seljak00,peacock00,cooray02}, which enable us to populate
haloes with galaxy in a way that accurately reproduces the galaxy
clustering. We use this technique on the Millennium simulation to
build galaxy catalogues mimicking Sloan Digital Sky Survey DR7
\citep[SDSS,][]{Abazajian09} volume-limited samples at
$z\simeq0.1$. We create two absolute magnitude-selected samples
corresponding to $M_r-5\log(h)<-18$ and $M_r-5\log(h)<-19$ from the
halo occupation measurements performed by \citet{zehavi11}. We choose
these cuts because they involve a significant fraction of the galaxies
residing in the low-mass end of the halo mass function. In practice to
create the galaxy catalogues, we populate haloes with central and
satellite galaxies using the mean occupation numbers given by the
HOD. While central galaxies are placed at halo centres, satellite
galaxies are randomly disposed around halo centres in such a way that
their radial distribution follows a NFW \citep{navarro96} density
profile. The details of the procedure are given in Appendix B of
\citet{delatorre12}. For each volume-limited sample we construct three
catalogues: one based on the original complete halo catalogue to which
we refer in the following as the fiducial sample; a second built from
a reconstructed halo catalogue below $m_{\rm lim}=10^{11.5}\mhmsun$
using $G=2.5\mhmpc$; and a third one built from a reconstructed halo
catalogue below $m_{\rm lim}=10^{11}\mhmsun$ using $G=1\mhmpc$. In
these reconstructions, we estimated the halo density field using the
DT method and assumed the power-law bias model in the conditional mass
function.

\begin{figure}
\centering
\includegraphics[width=88mm]{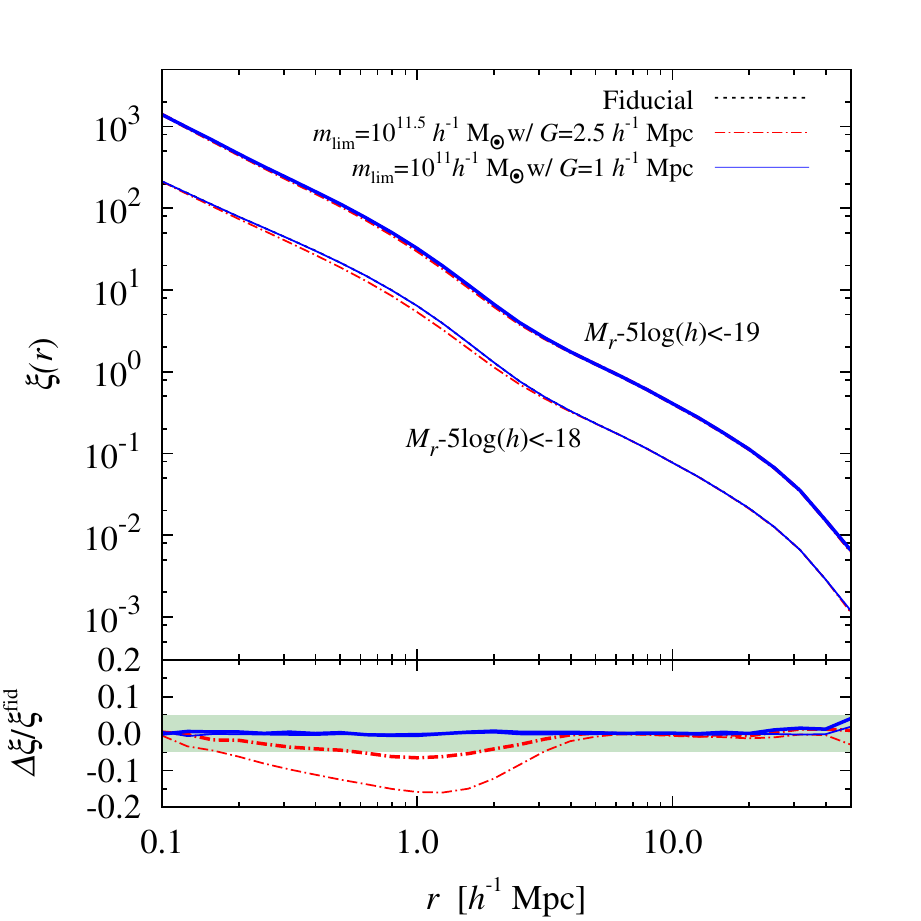}
\caption{Top panel: comparison of the two-point correlation functions
  of mock galaxies with $M_r-5\log(h)<-18$ (thin lines) and
  $M_r-5\log(h)<-19$ (thick lines) obtained from the original halo
  catalogue (referred as fiducial in the caption) with those obtained
  after reconstruction of haloes below $10^{11.5}\mhmsun$ (dot-dashed
  lines) and $10^{11}\mhmsun$ (solid lines) in the Millennium
  simulation. The two reconstructions have been performed respectively
  on grid sizes of $G=2.5\mhmpc$ and $G=1\mhmpc$. The amplitude of the
  correlation functions for the $M_r-5\log(h)<-18$ samples have been
  divided by $5$ to improve the clarity of the figure. Bottom panel:
  relative difference of two-point correlation functions of samples
  with reconstructed haloes with respect to that of fiducial
  samples. In the two panels the dotted and solid curves are not
  distinguishable as they almost overlap completely.}
\label{fig3}
\end{figure}

We present in Fig. \ref{fig3} the galaxy two-point correlation
functions for the two absolute magnitude-selected mock samples built
from the original complete halo catalogue. These are compared to those
measured in the mock samples in which the haloes of mass below $m_{\rm
  lim}=10^{11.5}\mhmsun$ and $m_{\rm lim}=10^{11}\mhmsun$ have been
reconstructed. In the case of the reconstruction with $m_{\rm
  lim}=10^{11.5}\mhmsun$ and $G=2.5\mhmpc$, we find that the
correlation functions obtained are well recovered, although the
correlation function is underestimated by up to $15\%$ on intermediate
scales for the $M_r-5\log(h)<-18$ sample. This underestimation is a
direct consequence of the resolution scale chosen for the
reconstruction. Indeed, the clustering drops on smaller scales than
the reconstruction scale for the reconstructed low-mass haloes. This
in turn causes the observed underestimation of the correlation
function of less luminous galaxies. However, by reconstructing the
halo density field on smaller scales, i.e. by using a lower mass limit
for the reconstruction of $m_{\rm lim}=10^{11}\mhmsun$, one can better
reproduce the halo clustering on $1\mhmpc$ scales and eliminate the
underestimation as shown in Fig. \ref{fig3}. The differences seen are
lower than one percent, below the impact of the simplifying assumption
of spherical haloes used in the HOD approach \citep{vandaalen12}. We
note that the small-scale galaxy clustering, i.e. below $0.7-1\mhmpc$,
is always well recovered. This comes from the fact that the galaxy
distribution inside haloes is independent of the halo clustering by
construction \citep{delatorre12} and the clustering on those scales is
dominated by this 1-halo term.

One could improve the method in the case of relatively coarse grid
reconstructions, by working at the sub-grid level and using additional
constraints from the mass non-linear power spectrum. Indeed, one could
envisage distributing haloes in each sub-cell so as to reproduce the
non-linear correlation function predicted from theory, instead of
randomly distributing them. We plan to investigate this extension of
the method elsewhere.

\section{Summary and conclusions}

We have described in this paper a method for populating dark matter
simulations with haloes of mass below the resolution limit. It is
based on estimating a continuous halo density field and then sampling
this stochastically in order to obtain low-mass haloes with the
correct abundance and bias. This latter part requires the conditional
halo mass function, which is extrapolated from the simulation itself
or taken from theoretical predictions.

We found that the method works well and allows us to reproduce
the halo distribution below the resolution limit with high fidelity,
in particular for reconstructions on grids of size $G=1\mhmpc$ or
below.  Moreover, the method is particularly efficient at producing
galaxy mock samples from low-resolution simulation halo catalogues. We
built galaxy mock samples using the HOD technique on the reconstructed
halo density field, tuned to mimic SDSS observations. We showed that
within a reasonable resolution limit range, one can recover the
overall two-point correlation function at the percent level.  

The method presented here is
relatively general. Another possible application is the reconstruction
of dark matter distributions. As with the galaxy mock sample
construction one can make use of the halo model to distribute dark
matter inside the haloes. Such datasets could then be used
to create cosmic shear catalogues from ray-tracing through the
simulation, or to predict the galaxy-lensing signal, a quantity
directly related to the galaxy-mass correlation function. Other
applications can readily be envisaged -- although we caution that
they should all be subject to the kind of validation tests
that we have performed here.

For the present, this validation has only been performed at
the two-point level; but with this restriction our results
demonstrate that one can communicate efficiently the full information
content of a large simulation by using only a catalogue of the more
massive haloes. This method should be very useful in the future in
building realistic galaxy mocks for the massive forthcoming
cosmological surveys such as Euclid \citep{laureijs11}, where the
volume and mass resolution requirements for the survey simulations are
both very high.

\section*{Acknowledgements}
We thank Gabriella De Lucia for giving us useful comments on the
manuscript. The MultiDark Database used in this paper and the web
application providing online access to it were constructed as part of
the activities of the German Astrophysical Virtual Observatory as
result of a collaboration between the Leibniz-Institute for
Astrophysics Potsdam (AIP) and the Spanish MultiDark Consolider
Project CSD2009-00064. The Bolshoi and MultiDark simulations were run
on the NASA's Pleiades supercomputer at the NASA Ames Research Center.

\setlength{\bibhang}{2.0em}
\setlength{\labelwidth}{0.0em}
\bibliography{biblio}

\bsp

\label{lastpage}

\end{document}